# Comparing Techniques for Aggregating Interrelated Replications in Software Engineering

Adrian Santos
M3S-ITEE University of Oulu, Finland
adrian.santos.parrilla@oulu.fi

Natalia Juristo
Universidad Politécnica de Madrid, Spain
natalia@fi.upm.es

## ABSTRACT

**Context:** Researchers from different groups and institutions are collaborating towards the construction of groups of interrelated replications. Applying unsuitable techniques to aggregate interrelated replications' results may impact the reliability of joint conclusions.

**Objectives:** Comparing the advantages and disadvantages of the techniques applied to aggregate interrelated replications' results in Software Engineering (SE).

**Method:** We conducted a literature review to identify the techniques applied to aggregate interrelated replications' results in SE. We analyze a prototypical group of interrelated replications in SE with the techniques that we identified. We check whether the advantages and disadvantages of each technique —according to mature experimental disciplines such as medicine— materialize in the SE context.

**Results:** Narrative synthesis and Aggregation of *p*-values do not take advantage of all the information contained within the raw-data for providing joint conclusions. Aggregated Data (AD) meta-analysis provides visual summaries of results and allows assessing experiment-level moderators. Individual Participant Data (IPD) meta-analysis allows interpreting results in natural units and assessing experiment-level and participant-level moderators.

**Conclusion:** All the information contained within the raw-data should be used to provide joint conclusions. AD and IPD, when used in tandem, seem suitable to analyze groups of interrelated replications in SE.

## 1 INTRODUCTION

Experimentation is common nowadays in SE [13, 23, 48]. However, two main threats to validity impact the suitability of individual experiments to generate reliable results[1]: the small sample sizes commonly gathered [13], and the lack of representativeness of the experimental settings to real-life contexts [48]. To overcome the previous weaknesses and to improve the generalizability of results, researchers from different groups and institutions are collaborating towards the construction of groups of interrelated replications [6, 17, 33, 38]. Collaborating with each other (e.g., by sharing the training material, cooperating during the design and execution of experiments, etc.), researchers are able to increase the sample size gathered at the overall level, and at the same time, check the effects of the technology under assessment in different settings. Eventually, this should improve the reliability of results and their generalizability towards different contexts [30]. These groups of replications (i.e., *groups of interrelated replications* or simply, *interrelated replications* from now onwards) provide certain advantages over groups of experiments gathered by means of Systematic Literature Reviews (SLRs) [26].

For example, researchers conducting interrelated replications have access to the raw-data of all the replications. This allows them to apply consistent analysis techniques and in turn, ensure that differences across replications' results do not emerge due to the different techniques applied to analyze each replication, but instead, due to real differences in the data gathered [9]. On the contrary, ensuring that differences across experiments' results are just due to the differences in the data gathered is unfeasible in SLRs, as in principle, different pre-processing or analysis techniques may have been followed to analyze each experiment.

Besides, as researchers conducting interrelated replications have first-hand knowledge of the settings of all the replications, they can ensure that, in case replications' results diverge, they are either due to genuine variability of results, or due to the specific changes introduced across replications —and thus, it is possible to elicit *moderators*, either at the *participant-level* (e.g., subjects' programming experience, etc.) or at the *experiment-level* (e.g., programming language). On the contrary, if experimental settings or populations are not fully described in research papers (what may happen due to length restrictions or reporting inconsistencies), this may limit SLRs' opportunities to identify moderators (as they may pass unnoticed), and at the same time, lower the reliability of SLRs' joint results or moderator effects (because of the potential existence of unacknowledged sources of variability impacting results [9]).

Even though meta-analysis of effect sizes is the *de facto* statistical method in SE to aggregate experiments' results in SLRs [26], several techniques (i.e., Narrative synthesis, Aggregated Data (AD) meta-analysis, Individual Participant Data (IPD) meta-analysis, and Aggregation of *p*-values) have been applied to aggregate interrelated replications' results according to a Systematic Mapping Study (SMS) that we conducted in 2017 [39]. Aggregating interrelated replications' results with unsuitable aggregation techniques may impact the reliability of the joint findings and eventually, to miss a valuable opportunity to obtain in-depth insights from experiments' results.

---

[1]Even though both *qualitative* results (e.g., text transcripts) and *quantitative* results (e.g., productivity scores in a percentage scale) can be collected in experiments, here we just focus on *quantitative* results, and the techniques that have been applied to aggregate them in SE.

In this paper we aim to answer a main **research question**:

- What are the advantages and disadvantages of the techniques used to aggregate interrelated replications' results in SE?

To answer this question, we first select a representative group of interrelated replications according to the results of our SMS [39]. Specifically, here we select a group of four interrelated replications on Test-Driven Development (TDD) with small and dissimilar sample sizes, heterogeneous results, different subject types (i.e., professionals and students), and identical experimental designs and response variable operationalizations. Then, we analyze our group of interrelated replications with all the aggregation techniques that we identified along our SMS [39]. Finally, we check whether the advantages and disadvantages of each technique —according to mature experimental disciplines such as medicine or pharmacology— materialize when analyzing our group of interrelated replications. Along this study we made several **findings**:

- Narrative synthesis and Aggregation of $p$-values fail to quantify the relevance of results, and do not take advantage of the full information contained within the raw-data to provide joint conclusions.
- IPD and AD complement to each other to aggregate interrelated replications' results and identify experiment-level moderators. However, IPD seems more suitable than AD to identify participant-level moderators.

In view of this, we suggest:

> IPD and AD seem suitable to be used in tandem to analyze groups of interrelated replications in SE.

This paper makes two main **contributions**:

- A *description and compilation of the advantages and disadvantages* of the aggregation techniques applied to analyze groups of interrelated replications in SE.
- *The first comparison* of the findings obtained on the same group of interrelated replications with the different aggregation techniques used in SE.

As a secondary contribution, we provide *a brief list of suggestions* to analyze groups of interrelated replications —based on the findings that we obtained from applying the techniques on a representative group of interrelated replications.

**Paper organization**. In Section 2 we provide the background of this study. In Section 3 we outline the research method followed. In Section 4 we provide a description of our group of interrelated replications. Next, in Section 5, Section 6, Section 7 and Section 8 we analyze our group of replications by means of Narrative synthesis, AD, IPD and Aggregation of $p$-values, respectively. In Section 9 we discuss the results achieved with each technique. Then, we outline the threats to validity of our study in Section 10. Finally, we provide the conclusions of this article in Section 11.

## 2 BACKGROUND

In this section, we first report on the importance of effect sizes and $p$-values to extract knowledge from interrelated replications' results (Section 2.1). Then, we provide an introduction to the aggregation techniques that have been used in SE to analyze groups of interrelated replications according to our SMS [39]. We go over them from the most, to the least used in SE: Narrative synthesis (Section 2.2), AD (Section 2.3), IPD (Section 2.4) and Aggregation of $p$-values (Section 2.5).

### 2.1 Effect Sizes and $p$-values

According to the latest recommendations provided by statistical reformers and associations [10, 32, 46], and those of interest groups on data analysis in medicine and pharmacology [29, 43], data analyses should quantify both the *practical significance* and the *statistical significance* of results.

Practical significance is usually quantified in terms of *effect sizes* and *95% confidence intervals* (i.e., *95% CIs*) [10, 46]. Effect sizes inform on the *magnitude* and *sign* of the relationship between two groups (or more generally, between two variables [7]). 95% CIs are commonly used as a quantifier on the precision of the effect size provided: the smaller the 95% CI, the larger the precision of the effect size and viceversa. Effect sizes can be conveyed in standardized units (i.e., units designed to rule out differences across experiments' response variables) or in unstandardized units (i.e., *natural* units). Examples of commonly used standardized effect sizes are Cohen's d and Hedge's g (i.e., Cohen's d small samples correction [4]). Commonly used unstandardized effect sizes are the parameter estimates of traditional statistical tests (e.g., the difference between the means of the two groups in a $t$-test, or the parameter estimates of linear regressions [4]).

Statistical significance is usually quantified in terms of *$p$-values*. $p$-values inform on the probability of obtaining the effect size observed —or a larger one— if there was no effect in the population (i.e., the effect size was equal to 0 in reality [8]). If the $p$-value is smaller than a certain threshold (e.g., 0.05 [8]), then results are declared as *statistically significant*. Put simply, if results are statistically significant, there is a genuine effect in the population according to the evidence collected (i.e., the raw-data). Unfortunately, one main shortcoming threatens the validity of $p$-values to inform on the relevance of results: $p$-values confound sample size and effect size [8]. Put differently, small $p$-values (and thus, significant results) may emerge not just because of the relevance of the effect size, but instead because of the presence of a large sample size —and perhaps a not so relevant effect size. Thus, some authors suggest replacing $p$-values by effects sizes and their corresponding 95% CIs instead [10, 32, 46], as they serve also to quantify the statistical significance of results (if the 95% CI does not cross 0, then the effect size is statistically significant), and they can be used either to quantify the precision of results.

In the following, we go over the aggregation techniques that have been used to analyze groups of interrelated replications in SE.

### 2.2 Narrative Synthesis

According to our SMS [39], Narrative synthesis was used to analyze 46% of the groups of interrelated replications. In Narrative synthesis, experiments' results —either in $p$-value or effect size terms— are "combined" together to provide a textual summary of results. For example, if Narrative synthesis is to be applied, first each experiment needs to be analyzed individually with a suitable statistical test (e.g., $t$-test, Wilcoxon test etc. [18]). Afterwards, all experiment' results are synthesized together, commonly following a "template" such as: *while results seem 'statistically significant/large/negative' in experiments X, Y and Z, results are not in experiment M. This difference of results could be due to H, N or K moderator variable.* The main advantage of Narrative synthesis is that it only needs experiment' effect sizes or *p*-values to provide joint results. Besides, Narrative synthesis allows to combine the results of experiments with wildly different designs and response variables into a joint conclusion (as it only requires either the *p*-values or the effect sizes [37]). However, Narrative synthesis fails to provide a quantitative summary of results (such as a joint effect size or a *p*-value [4]), and involves subjective judgment when providing joint results [35] (should all experiments be treated equally towards the joint conclusion? should experiments with practitioners be weighted more —as after all, they are more representative of reality?). Thus, different analysts may reach to disparate conclusions given the same raw-data [4].

## 2.3 AD

According to our SMS [39], AD was used to analyze 38% of the groups of interrelated replications. AD is commonly known as meta-analysis of effect sizes in SE [26]. To use AD, first all experiments' effect sizes need to be calculated —from either experiment-level summary statistics (e.g., mean, standard deviations, etc.), or from statistical tests' results (e.g., Pearson correlation [18]). Afterwards, effect sizes need to be combined into a joint effect size by means of a meta-analysis model. Two types of meta-analysis models can be fitted with AD [4]: fixed-effects models and random-effects models. Fixed-effects models rely on the assumption that differences across experiments' results are just due to random-sampling. Put differently, fixed-effects models assume that a *common underlying effect size* is being estimated across all the experiments. Random-effects models assume that differences across experiments' results are due to both random-sampling and real *heterogeneity of effects.* Put differently, instead of a common underlying effect size, in random-effects models, a *distribution of effect sizes* is being estimated. Generally speaking, in AD each experiment is *weighted* towards the overall result depending upon its sample size —if a fixed-effects model is used [4]— or depending upon its sample size and the total heterogeneity of results —if a random-effects model is used [4].[2] Finally, visual summaries of results such as forest-plots are usually provided to ease the understanding of results [4].

Figure 1 shows the corresponding forest-plot of an hypothetical group of four toy-replications.

At a simple glance at Figure 1, it can be seen that Experiment 3's results are remarkably different than those of Experiment 1, 2 and 4. Besides, the weight of each experiment towards the joint result (i.e., the stretched diamond at the bottom) can be assessed just by looking at the size of each black square (the larger the square, the larger the weight). In addition, the precision of the effect sizes can be simply ascertained by looking at their respective 95% CIs (the lines that cross the squares). Finally, heterogeneity of results can be assessed either visually (by observing the relative position of experiments' effect sizes and 95% CIs) or by means of the $I^2$ statistic or the *Q*-test: in this case, the heterogeneity of results is statistically

---
[2]Assuming a common variance term across experiments.

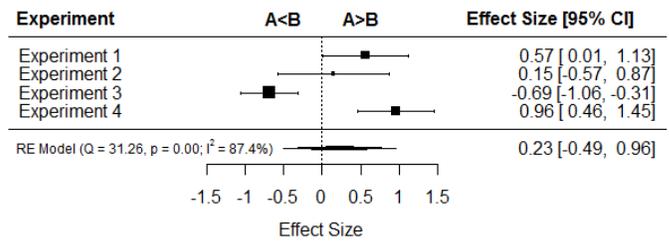

**Figure 1: Forest-plot for toy-replications.**

significant (i.e., *p*-value<0.001) and large (as $I^2 = 87.4$, $I^2$ is larger than 75% [4]).

If heterogeneity of results is identified with AD, "post-hoc" analyses should be conducted with the aim of eliciting moderator variables [4]. Sub-group meta-analyses and meta-regression (a generalization of sub-group meta-analysis) can be used for such ends [4]. Meta-regression is a special case of weighted regression where all experiments' effect sizes are regressed on either experiment-level moderators (e.g., programming language or testing tool), or on averaged participant-level moderators (e.g., participants' averaged programming experience in each experiment) [4].

Summarizing, the main advantages of AD are the appeal of forest-plots for summarizing results, and the availability of straightforward statistics for interpreting heterogeneity. AD's ability to combine the results of experiments with different response variable operationalizations and designs [4], and its suitability to assess experiment-level moderators [9, 21] are worth mentioning either. However, among AD's main shortcomings are its over-reliance on standardized effect sizes and thus, its lowered informativeness for conveying results (e.g., How *small* is a small Cohen's d of 0.23?). In addition, AD has a reduced statistical power for detecting participant-level moderators [28] and may also be affected by *ecological bias* (i.e., the presence of an averaged effect that may not be representative of the effect in the population) when identifying participant-level moderators [3].

## 2.4 IPD

IPD involves the central collection, processing and analysis of all experiments' raw-data into a joint conclusion [9]. IPD comes in two flavours: IPD mega-trial and IPD stratified [2]. According to the results of our SMS [39], IPD mega-trial and IPD stratified were used to analyze 33% and 15% of the groups of interrelated replications, respectively. In IPD mega-trial, the raw-data of all experiments are pooled together, and then analyzed as if data were coming from a single "big" experiment (e.g., by means of a *t*-test, or a U-Mann Whitney test [18]). IPD mega-trial may provide biased results if data are unbalanced across experiments and treatment groups [27] —what may happen in groups of interrelated replications with dissimilar sample sizes and missing data. IPD mega-trial may also have low statistical power if subjects resemble more to each other within experiments than across experiments [24] —what may happen in groups of interrelated replications with different types of subjects (e.g., professionals or students). Thus, IPD mega-trial shall be avoided by default [2, 16].

In IPD stratified, the raw-data of all experiments are analyzed jointly, but contrary to IPD mega-trial, in IPD stratified, the analysis is *stratified* by experiment -by including a factor accounting for "Experiment" within the statistical model fitted (e.g., an ANOVA model with the factors "Treatment" and "Experiment" [4]). IPD stratified models can be either fixed-effects models or random-effects models. Generally speaking, and as in AD, in IPD stratified models each experiment contributes towards the joint result depending upon its sample size[3]—if a fixed-effects model is used [16]— or depending upon its sample size and the total heterogeneity of results —if a random-effects model is used [16]. Typically used IPD fixed-effects models are ANOVA or ANCOVA —from the Generalized Linear Model family [18]. Typical IPD random-effects models are Linear Mixed Models (*LMM*) —from the Generalized Linear Mixed Models family [5].

For illustrative purposes, Table 1 shows the results of the group of four toy-replications presented in Section 2.3, analyzed now with a LMM.

**Table 1: LMM analysis for toy-replications.**

| Factor | Estimate | 95% CI | *p*-value |
|--------|----------|--------|-----------|
| Treatment A | 49.93 | (47.02, 52.85) | <0.001 |
| Treatment B | 51.82 | (41.66, 61.97) | <0.001 |
| $M_{Diff}$ | 1.88 | (-5.22, 8.99) | 0.601 |
| $sd_{Diff}$ | 6.71 | | |

As it can be seen in Table 1, the difference in performance between Treatment B ($M = 51.82$) and Treatment A ($M = 49.93$) is small (i.e., $M_{diff} = 1.88$). Besides, in comparison with the relatively small difference in performance, a large standard deviation of treatment effects is observed ($sd_{Diff} = 6.708$). Put differently, treatment effects span widely along a negligible mean effect. Thus, heterogeneity of treatment effects seems to have materialized.

If heterogeneity is detected, "post-hoc" analyses shall be conducted with the aim of eliciting moderator variables [47]. Statistical models with interaction terms between treatment and either experiment-level, or participant-level moderators shall be performed for such ends [20].

Among the main advantages of IPD are its increased statistical flexibility for accommodating missing-data (e.g., LMMs allow missing data —as long as data are missing at random [5]) and the possibility of interpreting results in natural units (instead of in standardized units as it is typical with AD). Besides, IPD allows assessing the effect of both experiment-level and participant-level moderators on results [20, 28]. However, among the main disadvantages of IPD are its reliance upon identical response variable scales across experiments [44], the unsuitability of some statistical models (e.g., repeated-measures ANOVAs) for analyzing together experiments with different designs [19], and the unavailability of straightforward statistics to interpret heterogeneity of results [5].

---

[3]Making the assumption that no interaction term between treatment and experiment is included within the statistical model, and that experiments share an identical variance term [16].

## 2.5 Aggregation of *p*-values

Aggregation of *p*-values was used to analyze just 7% of the groups of replications according to our SMS [39]. In Aggregation of *p*-values, each experiment is first analyzed by means of a suitable statistical model (e.g., such as a *t*-test, a Wilcoxon test etc. [19]), and then the *one-tailed p*-values of each experiment are combined into a joint *p*-value by means of either the Fisher's or Stouffer's method [4]. The main advantage of Aggregation of *p*-values is that it can combine the *p*-values of experiments with different designs or response variables into a joint result [35]. However, one of its main shortcomings is that Aggregation of *p*-values weights identically each experiment towards the joint conclusion regardless of their sample size, experimental design or quality [35] —even though more sophisticated versions of Aggregation of *p*-values also exist [47]. In addition, Aggregation of *p*-values does not provide a joint effect size, and thus, does not allow to quantify the relevance of results [4].

## 3 RESEARCH METHOD

We conducted a SMS in 2017 with the aim of identifying the techniques that had been used to aggregate interrelated replications' results in SE [39]. We also conducted a literature review to learn about the advantages and disadvantages of the aggregation techniques according to the literature of mature experimental disciplines such as medicine and pharmacology. Along our research, we came across numerous resources on meta-analysis [4, 47], statistical techniques to analyze multicenter clinical trials [11, 20, 41] and linear mixed models [5, 22].

After identifying all the techniques applied in SE, and learning about their advantages and disadvantages, in this study we wanted to assess whether such advantages and disadvantages would materialize in the SE context. For this, here we select a representative group of interrelated replications according to the results of our SMS [39]. Specifically, here we select a group of interrelated replications with the following characteristics: (1) four replications (i.e., the median number of replications within groups of replications) with small and dissimilar sample sizes; (2) an observable heterogeneity of results; (3) identical experimental designs and response variable operationalizations across replications (as in 85% and 77% of the groups of replications identified, respectively); (4) different subject types (as in 97% of the groups of replications).

Then, we analyze our group of replications with Narrative Synthesis, AD, IPD and Aggregation of *p*-values. We follow a similar Narrative synthesis procedure to that followed in the groups of replications that we identified along our SMS [39]. We follow Borenstein et al.'s procedures [4] for analyzing our group of replications with AD. We follow Whitehead et al.'s suggestions [47] and Fisher et al.'s recommendations [20] for analyzing our group of replications and identifying participant-level moderators, respectively. We use Fisher's method [4] to analyze our group of replications by means of Aggregation of *p*-values.

Once we analyze our group of replications with all the techniques, we compare the findings that we obtained with each one. Finally, we asses the extent to which the advantages and disadvantages of the techniques materialize in the SE context.

## 4 INTERRELATED REPLICATIONS ON TDD

A group of four interrelated replications has been conducted to evaluate the effects of TDD on quality. Three replications were conducted at F-Secure -a multi-national security and digital company- and one at UPV -Technical University of Valencia. The threats to validity of F-Secure's experimental design, experiment protocol, instruments and experimental tasks were published elsewhere [45]. UPV is a close replication of F-Secure.

### 4.1 Dependent and independent variables

The independent variable across all the replications is **development approach**, with TDD and Iterative-Test Last (ITL) as treatments. ITL is defined as the reverse-order approach of TDD following Erdogmus et al. [14].

The dependent variable across all the experiments is **functional correctness**. Functional correctness is a sub-characteristic of quality according to ISO 25010. It is defined as *the degree to which a system provides the correct results with the needed degree of precision* [1]. We measure functional correctness as the percentage of test cases that successfully pass from a battery of test cases that we built for testing participants' solutions. Specifically, we measured functional correctness as:

$$FC = \frac{\#Test\ Cases(Pass)}{\#Test\ Cases(All)} * 100$$

All replications have an identical *experimental design*: an AB within-subjects design [23] (i.e., a repeated measures design where each subject first applies ITL and then TDD).

### 4.2 Subjects

Subjects were handed a survey in each replication. The survey contained a series of ordinal-scale (i.e., inexperienced, novice, intermediate and expert) self-assessment questions with regard to their experience in programming, Java, unit testing and JUnit.[4]

**Table 2: Mean experiences across replications.**

| Experiment | N | Programming | Java | Unit | JUnit |
|---|---|---|---|---|---|
| F-Secure H | 6 | 3.67 | 2.33 | 2.17 | 2.17 |
| F-Secure K | 11 | 2.91 | 1.82 | 1.64 | 1.27 |
| F-Secure O | 7 | 3.29 | 2.71 | 2.71 | 2 |
| UPV | 33 | 2.36 | 1.88 | 1.04 | 1 |

Table 2 shows the mean experiences (Programming, Java, Unit and JUnit) of the participants' across replications (i.e., 1-4, for inexperienced, novice, intermediate and experts, respectively).[5] As it can be seen in Table 2, the most senior developers are those at F-Secure O and F-Secure H, while those at UPV are the ones with lowest experience. As a summary, **our group of replications is comprised by an heterogeneous population of TDD novices.**

---

[4]The survey and its results were published elsewhere [12].
[5]For simplicity's sake, we consider the ordinal variables along this study as continuous. This approach is commonly followed in other areas [34].

### 4.3 Descriptive Statistics

Table 3 shows the descriptive statistics (i.e., sample size, mean, standard deviations, and medians) for ITL and TDD's FC scores across all the replications.

**Table 3: Descriptive statistics: ITL vs. TDD.**

| Experiment | Treat. | N | Mean | SD | Median |
|---|---|---|---|---|---|
| F-Secure H | ITL | 6 | 30.71 | 36.58 | 24.16 |
| | TDD | 6 | 40.23 | 33.43 | 35.34 |
| F-Secure K | ITL | 11 | 22.17 | 20.44 | 17.98 |
| | TDD | 11 | 35.42 | 35.40 | 22.41 |
| F-Secure O | ITL | 7 | 16.05 | 20.81 | 7.87 |
| | TDD | 7 | 68.97 | 31.53 | 81.03 |
| UPV | ITL | 31 | 33.38 | 39.79 | 6.74 |
| | TDD | 29 | 77.16 | 21.04 | 83.93 |

Figure 2 shows the profile-plot of each experiment's ITL and TDD mean FC scores.

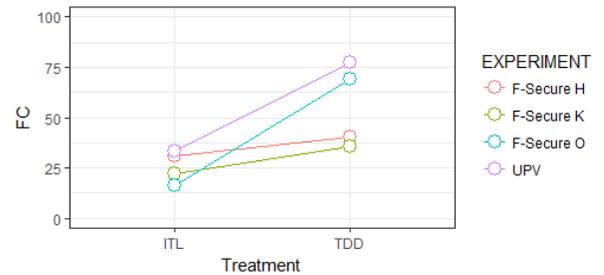

**Figure 2: Profile-plot: ITL vs. TDD.**

As it can be seen in Figure 2, the mean FC scores achieved with ITL are clumped together around 25, while the mean FC scores of TDD vary largely across experiments. Besides, the difference in performance between TDD and ITL (i.e., the slope of the lines) varies largely across experiments: while F-Secure K's line is almost flat (and thus, ITL behaves similarly to TDD), F-Secure O's line looks much steeper (TDD outperforms ITL to a large extent). At first glance, **heterogeneity of results seems to have materialized.**

Next, we analyze the group of replications with all the techniques that have been applied in SE [39].

## 5 NARRATIVE SYNTHESIS

To analyze a group of replications with Narrative synthesis, first, each replication needs to be analyzed individually. As all replications have an identical AB within-subjects design, we analyze each of them with a dependent $t$-test [19]. The dependent $t$-tests' results are shown in Table 4.

In Narrative synthesis, after analyzing each replication individually, the $p$-values —or the signs or magnitudes— of the effect sizes (this is, the treatment estimates of the dependent $t$-tests) are compared to provide a textual summary of results. For example: as it can be seen in Table 4, TDD seems to outperform ITL across all the replications (i.e., as all the effect sizes are positive). However, as just results are statistically significant at F-Secure O and UPV, and not att F-Secure H or F-Secure K (and thus, an identical number of experiments point in different directions), we cannot draw a definite conclusion about the significance of results. Finally, heterogeneity seems to have materialized: effect sizes vary largely across replications (i.e., from $M = 52.91$ at F-Secure O to $M = 9.52$ in F-Secure H). Thus, differences across replications' effect sizes may be due to experiment-level moderators (e.g., subject type), or due to participant-level moderators (i.e., participants' experiences). However, we cannot observe any pattern on results: TDD outperforms ITL to a large extent for both students (i.e., UPV) and professionals (F-Secure O). Besides, participants' mean experiences seem not much informative either (UPV's participants seem to outperform those of F-Secure H and F-Secure K, despite being novices). In view of this, what we conclude by means of Narrative synthesis is that more replications are needed to assess the significance of results and understand the circumstances that favour TDD over ITL.

**Table 4: Individual analyses: ITL vs. TDD.**

| Experiment | Estimate | 95% CI | *p*-value |
|---|---|---|---|
| F-Secure H | 9.52 | (-19.58, 38.62) | 0.483 |
| F-Secure K | 13.26 | (-7.26, 33.77) | 0.193 |
| F-Secure O | 52.91 | (30.44, 75.39) | **<0.001** |
| UPV | 42.31 | (29.02, 55.62) | **<0.001** |

## 6 AD

### 6.1 Main Analysis

To analyze a group of replications by means of AD, first, all replications' effect sizes —and their corresponding standard errors— need to be computed. Then, all replications' effect sizes need to be combined by means of a meta-analysis model. Finally, results are usually represented in a forest-plot.

First, we calculate all replications' Hedge's g —and their corresponding standard errors— by means of summary statistics (i.e., mean, standard deviations, sample size per treatment group, and correlations between ITL and TDD's FC scores [4]). As five subjects at UPV had missing data, their data had to be removed to calculate the effect size (as they could not contribute towards the estimation of the correlation term required by the standard error formulae [4]). Afterwards, we fitted a random-effects model to pool all the effect sizes together. Figure 3 shows the forest-plot corresponding to analyzing the group of replications with AD.

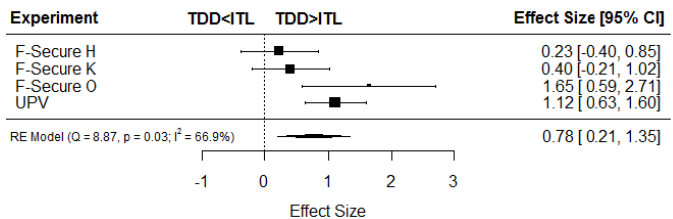

**Figure 3: Forest-plot: ITL vs. TDD.**

As it can be seen in Figure 3, F-Secure O and UPV effect sizes are large (i.e., larger than 0.8 [4]), and statistically significant (i.e., their confidence intervals do not cross 0). However, F-Secure H and F-Secure K effect sizes are small (i.e., lower than 0.5 [4]). Combining all effect sizes together (see the stretched black-diamond at the bottom), TDD outperforms ITL to an —almost— large extent ($M = 0.78; 95\% \; CI = (0.21, 1.35)$). However, an observable and statistically significant heterogeneity of results materialized ($I^2 = 66.9\%, Q = 8.87; p = 0.03$). Thus, moderator variables —either at the experiment-level or at the participant-level— shall be uncovered with AD with the aim of explaining the differences across replications' results.

### 6.2 Moderators Analysis

Sub-group meta-analyses are usually performed with AD to identify experiment-level moderators. Table 5 shows the results of performing a sub-group meta-analysis on the influence of subject type (i.e., professionals vs. students) on results.

**Table 5: Sub-group meta-analysis: experiment-level moderator.**

| Variable | Group | N | Estimate | 95% CI | $I^2$ |
|---|---|---|---|---|---|
| Subject | Professionals | 3 | 0.65 | (-0.10, 1.41) | 68.07% |
| | Students | 1 | 1.12 | (0.63, 1.60) | 0 |
| | Difference | - | 0.47 | (-0.44, 1.36) | - |

As it can be seen in Table 5, students obtain more benefit than professionals with TDD ($M = 1.12$ and $M = 0.65$ for students and professionals, respectively). Even though the difference in performance between students and professionals is noticeable ($M = 0.47$), this is not statistically significant (i.e., the 95% CI of the difference crosses 0). As a summary, and in view that relevant differences across sub-groups are not statistically significant, we conclude, our group of replications seems under-powered for uncovering experiment-level moderators with AD.

Finally, with the aim of identifying participant-level moderators, we performed a meta-regression with each of the experience variables that we measured along the survey (experience with programming, Java, unit testing and JUnit). In meta-regression, the average experiences of all the subjects within each experiment (X-axis) are regressed against the effect sizes of the experiments (Y-axis). Figure 4 shows the results of the meta-regressions that we performed. Table 6 shows the estimates of the interaction terms according to meta-regression.

**Table 6: Meta-regression: participant-level moderators.**

| Interaction | Estimate | 95% CI | *p*-value |
|---|---|---|---|
| Programming | -0.4 | (-1.68, 0.84) | 0.51 |
| Java | 0.61 | (-1.38, 2.60) | 0.55 |
| Unit testing | 0.12 | (-1.08, 1.32) | 0.84 |
| JUnit | -0.16 | (-1.60, 1.28) | 0.82 |

As it can be seen in Figure 4, the performance with TDD decreases the larger the average experience with programming or JUnit (as it can be seen by the negative slope of the lines). Besides, while the experience with unit testing seems to have almost no impact on results (i.e., the line is almost flat along all the experience levels), the larger the experience with Java, the larger the benefit with TDD. Thus, we hypothesize, subjects knowledgeable in Java may focus more on the TDD process itself and less on syntax-related problems while coding. This way, they manage to obtain larger quality scores than those without any previous experience with Java. However, as the experience with Java is not statistically significant, we conclude that further experiments shall be conducted to elicit participant-level moderators with AD.

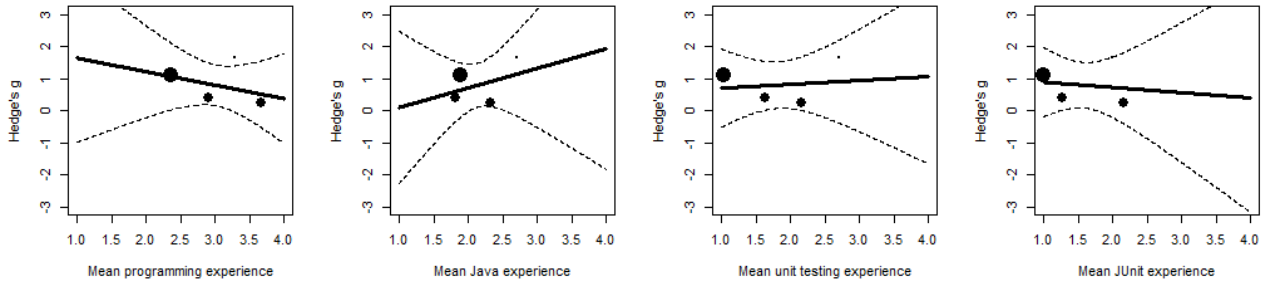

**Figure 4: Meta-regression plot: participant-level moderators.**

## 7 IPD

### 7.1 Main Analysis

To analyze a group of interrelated replications with IPD, the raw-data of all replications need to be analyzed together by means of a statistical test. Table 7 shows the results of analyzing our group of replications with a LMM.

**Table 7: LMM analysis for the group of replications.**

| Factor | Estimate | 95% CI | *p*-value |
|---|---|---|---|
| ITL | 27.44 | (13.08, 41.79) | <0.001 |
| TDD | 56.27 | (22.12, 90.42) | <0.001 |
| $M_{Diff}$ | 28.83 | (9.72, 47.93) | 0.004 |
| $sd_{Diff}$ | 16.09 | | |

As it can be seen in Table 7, the difference in performance between TDD ($M = 56.27$) and ITL ($M = 27.44$) is large ($M_{Diff} = 28.83$) —at least compared to the mean of the ITL group. Put differently, TDD seems to double the performance of ITL. Besides, the standard deviation of the differences between TDD and ITL ($sd_{Diff} = 16.09$) is large when compared to the difference between the means (16.09/28.75, around a 60% variation around the mean). Thus, an observable heterogeneity of treatment effects materialized in the group of replications. Moderator variables shall be identified with IPD to explain the heterogeneity of results.

### 7.2 Moderators Analysis

To assess moderator effects with IPD, statistical models with interaction terms need to be fitted [20]. As previously done with AD,

here we first assess the effects of subject type (i.e., professionals vs. students) on results. To do so, we fit an LMM with the interaction term between treatment and subject type. Table 8 shows the results of the interaction term in the LMM fitted.

**Table 8: LMM results: experiment-level moderators.**

| Interaction | Estimate | 95% CI | *p*-value |
|---|---|---|---|
| Subject:Students | 16.32 | (-37.16, 69.55) | 0.545 |

As it can be seen in Table 8, the difference in performance between students and professionals with TDD seems relevant ($M = 16.32$), at least when compared to the difference in performance between TDD and ITL previously identified without the interaction term ($M = 28.83$). In other words, students seem to obtain an increase of around a 60% in FC scores (16.32/28.83) when compared to professionals. However, the difference in performance between students and professionals is not statistically significant. Thus, according to IPD, the group of replications seems under-powered for detecting experiment-level moderators.

Finally, we assess the effect of participants' experiences on results by fitting four different LMMs (one per experience variable). The parameter estimates of the interaction terms are presented in Table 9. Figure 5 shows the corresponding regression lines of the interaction terms.

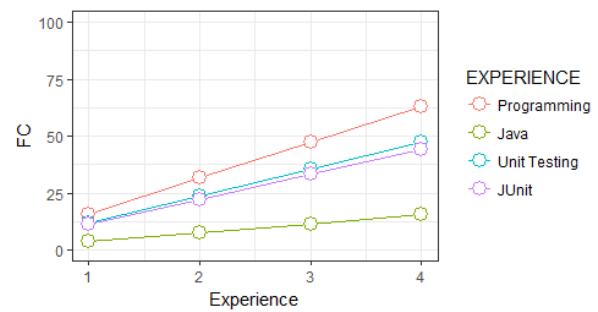

**Figure 5: LMM interactions: participant-level moderators.**

As it can be seen in Figure 5, the larger the subjects' experiences with either programming, Java, unit testing or JUnit, the larger the difference in performance between TDD and ITL. For example, while the differences between TDD and ITL in mean FC scores among novice programmers are relatively small (see the red line when Experience equals to 1), the difference in performance increases to a considerable extent when subjects are very experienced (up to a difference of around 60 points in FC scores). Besides, the effect of programming experience seems relevant ($M = 15.76$) and statistically significant. As a conclusion, according to IPD, the larger the experience with programming, the larger the benefit with TDD in comparison with ITL. In addition, according to IPD, we achieved enough statistical power to detect participant-level moderators within the group of replications.

**Table 9: LMM: participant-level moderators.**

| Interaction | Estimate | 95% CI | $p$-value |
|---|---|---|---|
| Programming | 15.76 | (0.49, 31.04) | 0.04 |
| Java | 3.85 | (-8.13, 15.83) | 0.52 |
| Unit testing | 11.79 | (-5.95, 29.54) | 0.18 |
| JUnit | 11.07 | (-6.26, 28.41) | 0.20 |

## 8 AGGREGATION OF $p$-VALUES

To analyze a group of replications by means of Aggregation of $p$-values, first, each experiment has to be analyzed individually with a statistical test (e.g., by means of the dependent $t$-test). Afterwards, the *one-tailed* $p$-values need to be combined by means of the Fisher's method. Thus, to analyze our group of replications, we first analyze each replication individually with a *one-tailed* dependent $t$-test and then we combine all the $p$-values by means of the Fisher's method. A statistically significant difference between TDD and ITL was obtained ($\tilde{\chi}^2$=47.13; $df$=8; $p$<0.001). Thus, by means of Aggregation of $p$-values, we conclude that in at *least one experiment*, TDD outperforms ITL.

## 9 DISCUSSION

Table 10 shows a summary of the findings that we made with each technique in our group of replications.

As it can be seen in Table 10, we obtained different insights with each technique. Besides, different advantages and disadvantages materialized with each technique. For example, with Narrative synthesis it was possible providing joint results in a straightforward fashion without worrying about replications' designs or response variables' scales (as just a textual summary of results had to be provided). However, it was unfeasible providing any quantitative summary of results (e.g., an effect size or $p$-value). Thus, in our group of replications we could not quantify the extent to which TDD outperformed ITL at the group level. Besides, contradicting results emerged in our group of replications: two replications provided non-significant results and two others provided significant results. Shall these be considered as contradicting evidence? Or instead, shall industrial experiments' results overtake those of academic experiments? After all, industrial experiments may be more representative of reality. Thus, subjective judgment may affect the conclusions reached with Narrative synthesis. Finally, one last shortcoming of Narrative synthesis emerged: we could not find any

"pattern" on results, and thus, we were not able to identify moderators that may explain why TDD outperformed ITL that differently across experiments.

Contrary to Narrative synthesis, AD allowed to provide a joint effect size and $p$-value: TDD outperformed ITL to an almost *large* (according to Hedge's g rules of thumb [4]) and *statistically significant* extent. Besides, AD allowed to visualize the extent to which TDD outperformed ITL across the replications (by simply looking at the forest-plot). AD also allowed to quantify the heterogeneity of results in easy-to-interpret statistics such as the $I^2$: an observable and almost large heterogeneity of results emerged. Finally, it was possible evaluating the extent to which experiment-level moderators affected results: a *small* —but close to medium— and not *statistically significant* difference in performance between students and professionals was identified. However, we could not interpret results in a straightforward fashion: *How* small was the effect of subject type? *How* large were joint results? Even though rules of thumb exist for interpreting results, we could not convey results in natural units of difference between the performance of TDD and ITL. Besides, ecological bias emerged while eliciting participant-level moderators: experiment-averaged programming experience seemed to influence negatively to the performance achieved with TDD (the larger the experience, the smaller the effect size). This result was the opposite of what we found with IPD. This misleading conclusion emerged with AD as one of the experiments with the larger effect sizes (i.e., UPV) was comprised by subjects with a low experience on-average, and besides, experiments with lower effect sizes (i.e., F-Secure H or F-Secure K) were comprised by subjects with larger experiences on-average. Thus, in view of the evidence available at the experiment-level, AD fitted a meta-regression line with a negative relationship between effect size and average programming experience. Finally, neither experiment-level nor participant-level moderators could be identified: *our group of replications seemed under-powered for detecting moderators according to AD.*

As with AD, with IPD it was possible providing a joint effect size (i.e., treatment estimate) and a $p$-value: the difference in performance between TDD and ITL was relevant and statistically significant. However, contrary to AD, with IPD it was possible interpreting joint results in natural units: *subjects doubled their performance with TDD compared to ITL* (as the mean of the TDD group was twice as large as that of the ITL group). IPD offered an extra advantage over AD: subjects with missing data (i.e., 5 subjects in the UPV experiment) could still be analyzed with IPD —as a LMM was used [5]— contrary to AD, where their data had to be discarded to calculate the standard error of the Hedge's g —as the data of such subjects could not contribute towards the calculation of the correlation between the TDD and ITL groups [4]. Despite we experienced some of the advantages of IPD, we also experienced some of its shortcomings: the assessment of heterogeneity was not straightforward (as the standard deviation of the differences between TDD and ITL's means had to be compared with the mean effect). Where is the limit to claim heterogeneous results? Finally, even though IPD and AD seemed equally suitable to assess the influence of subject type of results, IPD was superior to AD to identify participant-level moderators: a relevant and statistically significant influence of programming experience on TDD's performance was identified with IPD. Thus, *according to IPD our group of replications was adequately powered to identify participant-level moderators.*

**Table 10: Findings with each technique in our group of interrelated replications.**

| Technique | Result | Effect Size | p-value | Moderators | Comments |
|---|---|---|---|---|---|
| Narrative | TDD>ITL | ? | ? | ? | Conflicting p-values. No patterns in moderators. |
| AD | TDD>ITL | Large | ✓ | *Under-powered* | Standardized units. Ecological bias. |
| IPD | TDD>ITL | Large | ✓ | Programming experience | |
| Agg. p-values | TDD>ITL | ? | ✓ | ? | Significance was already known. |

Finally, Aggregation of *p*-values provided a joint *p*-value as conclusion. As the *p*-value was statistically significant, *at least the results of one replication are statistically significant*. However, that was already known before aggregating results (as UPV and F-Secure H results were already significant). Besides, no effect size was provided by Aggregation of *p*-values. This impacts the interpretation of results: how relevant was the overall effect?

To wrap-up, and based on our findings when applying the techniques to a representative group of replications in SE, we provide a *series of recommendations*:

- AD and IPD seem more suitable than Narrative synthesis and Aggregation of *p*-values to analyze groups of interrelated replications, specially for identifying moderators.
- AD and IPD seem to complement to each other for aggregating experiments' results and assessing experiment-level moderators.
- If all replications have identical response variable operationalizations, there is no need to use standardized units with AD. Go ahead and use AD with natural units or use IPD instead. This may increase the interpretability of results.
- Groups of interrelated replications may be under-powered to detect experiment-level moderators. The possibility of assessing participant-level moderators can be seen as a strength of groups of interrelated replications. IPD seems suitable to identify participant-level moderators.

## 10 THREATS TO VALIDITY

*One group of replications, can we generalize?* Due to length restrictions we could just apply all the aggregation techniques on a sole group of interrelated replications. However, we selected a representative group of interrelated replications according to the results of our SMS [39]. Even though our findings are not conclusive, they seem to agree with those achieved in medicine and pharmacology. Specifically, IPD seems to provide similar results as AD for assessing treatment effects or experiment-level moderators [9, 21], and IPD seems superior to AD for assessing participant-level moderators [20, 28].

*One statistical method per aggregation technique, how limited are the findings?* Due to length restrictions we could just use a single statistical method (i.e., *t*-test, random-effects meta-analysis, linear mixed model, and Fisher's method) per aggregation technique (i.e., Narrative synthesis, AD, IPD and Aggregation of *p*-values, respectively). Even though we acknowledge this limitation, we tried to use commonly used statistical tests in SE (i.e., the *t*-test [13]), random-effects meta-analysis models and Linear Mixed Models as they are the ones that shall be used whenever heterogeneity materializes [4, 5] —as it is commonly the case in SE experiments [36, 42]— and Fisher's method as it is the Aggregation of *p*-values method most used to aggregate interrelated replications' results in SE [39].

*Parametric tests and effect sizes, are there any threats to the analysis approach?* Parametric statistical tests and effect sizes (e.g., *t*-test, LMMs, Cohen's d, etc.) may be unsuitable to analyze non-normal data [48]. However, along this study we relied upon parametric statistical tests as they are robust to departures from normality, even in smaller data-sets than those typical in SE experiments [15, 34]. Besides, the larger the sample size —as it happens when pooling together the raw-data of all replications— the larger their robustness to departures from normality [31]. We relied upon parametric effect sizes as they are by far, the most used in SE [25].

*One analyst, how biased are results?* The measurement of the adequacy of each technique to analyze groups of replications may be sensitive to plausible bias due to researchers' preferences. In order to address this issue, we adopted similar procedures to those followed in medicine and pharmacology to provide joint conclusions [4, 47], and to assess experiment-level and participant-level moderators with either AD [4, 40] and IPD [20]. Researchers were triangulated to counteract the subjectivity of the main analyst (i.e., the first author) during the data analysis phase.

## 11 CONCLUSION

Narrative synthesis, Aggregation of *p*-values, Aggregated Data (AD) and Individual Participant Data (IPD) have been used to analyze groups of interrelated replications in SE [39]. If access to the raw-data is guaranteed, all the information contained within the raw-data should be used to provide joint conclusions.

Narrative synthesis does not provide a quantitative summary of results and involves subjective judgment when yielding joint conclusions or identifying moderator variables. Aggregation of *p*-values fails to quantify the relevance of results and weights identically each experiment towards the joint conclusion. AD and IPD seem suitable to analyze groups of replications and assess experiment-level moderators. AD and IPD allow providing quantitative summaries of results, weight transparently each experiment towards the joint conclusion, and in case results differ, assess the effect of moderators on results. AD seems suitable because of its intuitiveness and appealing visual summaries. IPD should seems suitable because of its adequacy for handling missing data and offering joint results in natural units. The possibility to assess participant-level moderators is an advantage in groups of interrelated replications (as the raw-data are available). IPD seems superior to AD for this.

## ACKNOWLEDGMENTS

This research was developed with the support of the Spanish Ministry of Science and Innovation project TIN2014-60490-P.